# Resolution of the Klein paradox

A. D. Alhaidari

*Saudi Center for Theoretical Physics, Jeddah, Saudi Arabia*

We present a resolution of the Klein paradox within the framework of one-particle relativistic quantum mechanics. Not only reflection becomes total but the vacuum remains neutral as well. This is accomplished by replacing the pair production process with virtual negative energy "incidence" within the barrier in a similar manner to what is done with image charges in electrostatic and virtual sources in optics.

The physics and mathematics of the Dirac equation is very rich and illuminating. However, this is true only if one includes the complete solution space of the equation. It is well known that the Dirac equation has positive as well as negative energy solutions [1]. The negative energy solutions are the subject of various interpretations that wax and wane throughout the history of the equation. However, since the equation is linear, then the complete solution must be a linear combination of the two. Physical and mathematical results and interpretations thereof are correct only if the full contribution of the complete solution is accounted for. Klein's paradox [2] results from the conventional solution of the Dirac equation for a potential step of height $V$ that is larger than $2m$, where $m$ is the rest mass of the particle ($\hbar = c = 1$). If the energy of the particle is in the range $+m < E < V-m$ (known as the Klein energy zone), then partial reflection not total reflection will result despite the fact that the energy is lower than the height of the barrier. The traditional resolution of the paradox uses tools outside one-particle relativistic quantum mechanics where particle/anti-particle pair creation is employed. In this work, we show that the paradox could be resolved within quantum mechanics by replacing the *physical* pair production process by *virtual* negative energy scattering under the barrier. We gauge these negative energy solutions with our own interpretation in which we follow a procedure similar to that in optics where virtual sources are included in the unphysical region to obtain the correct solution in the physical region. Another interpretation of these virtual negative energy solutions is also found in electrostatics where virtual image charges are added to obtain the correct solution.

Figure 1 shows the configuration associated with the Klein paradox. The typical interpretation is that a beam of electrons with energy in the range $[+m, V-m]$ incident from left gets partially reflect at the barrier [3]. To account for the electrons transmitted into the barrier, concepts like "charged vacuum" and spontaneously produced electron-positron pairs …etc, come into play. In most of earlier attempts at resolving the paradox, principles and tools like these, which come from fields outside relativistic quantum mechanics (e.g. second quantization, quantum field theory, many particle physics, …etc.) are deployed. In our view, however, a successful resolution of the paradox must come about from within the framework of one-particle relativistic quantum mechanics where the paradox was originally posed. The assertion that the theory at strong coupling is not sufficient to describe this scattering process is debatable since the theory is found to be correct even at higher energies where $E > V - m$ [1]. To address the paradox properly, it helps first to look at the square barrier problem with a particle beam incident from left (See Fig. 2). The solution of the wave equation to the right of the square barrier consists of positive energy plane waves traveling in the $\pm x$ directions. However, the physical boundary conditions allow only for transmitted waves to the right. On the other hand, in the potential step problem the solution of the Dirac equation to the right of the barrier ($x > 0$) in the Klein energy zone consists of negative energy plane waves traveling in the $\pm x$ directions. Here, however, we will not dismiss the plane wave solutions of negative energy anti-electrons traveling to the left and incident on the barrier. We will utilize this part, which is usually missing from the conventional solution, in the resolution of the paradox and we give it a proper interpretation. We start by making analogy with the electrostatic and optical model.

In the electrostatic problem (Fig. 3a), positive and negative charges are induced at the boundary (the surface at $x = 0$). Those with the same sign as the source charge will be displaced to $+\infty$ whereas those with the opposite sign will remain at the boundary. The latter are then replaced by a fictitious negative image charge (Fig. 3b). Due to the linearity of the problem, the correct solution (e.g., the electric field) in the physical region, $x < 0$, is obtained as superposition of those associated with the *real* and *fictitious* charge. The process of charge induction at the boundary is analogous to the process of pair creation at the potential barrier. Consequently, we will develop a virtual process in the atomic system that mimics the virtual image charge for obtaining the correct solution. In the optical model, on the other hand, full reflection of light from a light source on front of the plane mirror (left region in Fig. 4a) could be replicated by replacing the mirror (solid vertical line in Fig. 4a) with a partially transmitting plane glass (broken vertical line in Fig. 4b) and placing an identical source of light behind the glass in the unphysical region (see Fig. 4b). The figure shows the optical beams that correspond to the reflected ($R$ and $\mathcal{R}$) and transmitted ($T$ and $\mathcal{T}$) electronic beams. It is not difficult to see that $|R|^2 + |\mathcal{T}|^2 = 1$. The following is a brief technical presentation of the resolution of the paradox within relativistic quantum mechanics (without resorting to pair creation) that utilizes virtual negative energy incidence in the region $x > 0$ and in a manner similar to those in the electrostatic and optical models.

In the conventional relativistic units, $\hbar = c = 1$, the steady state Dirac equation for this 1D problem could be written as follows [1,4]

$$\begin{pmatrix} m + V(x) - E & -\frac{d}{dx} \\ \frac{d}{dx} & -m + V(x) - E \end{pmatrix} \begin{pmatrix} \psi^+(x) \\ \psi^-(x) \end{pmatrix} = 0. \tag{1}$$

The potential enters in the equation as the time component of a vector with vanishing space component. For $x < 0$, where $V(x) = 0$, this equation relates the two spinor components as follows

$$\psi^{\mp}(x) = \frac{1}{m \pm E} \frac{d}{dx} \psi^{\pm}(x), \tag{2}$$

which is NOT valid for $E = \mp m$. We also obtain the following Schrödinger-like second order differential equation

$$\left( \frac{d}{dx^2} + E^2 - m^2 \right) \psi^{\pm}(x) = 0. \tag{3}$$

Now, since $E = \mp m$ belongs to the negative/positive energy spectrum, then Eq. (2) and Eq. (3) with the top/bottom sign are valid ONLY for positive/negative energy, respectively. We should emphasize that Eq. (3) does NOT give the two components of the spinor that belong to the SAME energy subspace. One has to choose one sign in Eq. (3) to obtain ONLY one of the two components then substitute that into Eq. (2) with the corresponding sign to obtain the other component. The positive and negative energy subspaces are completely disconnected. This is a general feature of the solution space of the Dirac equation, which is also sometimes overlooked. Now, for $x > 0$, the same analysis follows but with the substitution $E \to E - V$.

We start by giving the traditional solution of the problem [3]. The positive energy spinor wavefunction in the Klein energy zone ($m < E < V - m$) for $x < 0$ is

$$\psi(x) = \frac{1}{\sqrt{1+\alpha^2}} \begin{pmatrix} 1 \\ i\alpha \end{pmatrix} e^{ikx} + \frac{A}{\sqrt{1+\alpha^2}} \begin{pmatrix} 1 \\ -i\alpha \end{pmatrix} e^{-ikx}, \quad (4)$$

where $k = \sqrt{E^2 - m^2}$ and $\alpha = \sqrt{(E-m)/(E+m)}$. This solution represents two positive energy electron beams. One beam of unit amplitude incident from left and another reflected beam of amplitude $A$. For $x > 0$, the negative energy solution representing the transmitted beam is

$$\psi(x) = \frac{B}{\sqrt{1+\beta^2}} \begin{pmatrix} -i\beta \\ 1 \end{pmatrix} e^{-ipx}, \quad (5)$$

where $p = \sqrt{(V-E)^2 - m^2}$ and $\beta = \sqrt{(V-E-m)/(V-E+m)}$. One should note that for positive (negative) energy, $e^{\pm iqx}$ is a wave traveling in the $\pm x$ ($\mp x$) direction, respectively, where $q$ is the positive wave number or linear momentum. Matching the spinor wavefunction at $x = 0$ gives

$$A = (\alpha\beta - 1)/(\alpha\beta + 1), \quad B = \frac{2i\alpha}{\alpha\beta+1} \sqrt{\frac{1+\beta^2}{1+\alpha^2}}. \quad (6)$$

The traditional solution stops here with the interpretation that the reflection amplitude is $R = A$ and the transmission amplitude is $T = B\sqrt{\frac{\beta}{\alpha}\frac{1+\alpha^2}{1+\beta^2}}$. Of course, reality gives $|T|^2 + |R|^2 = 1$. However, the resulting reflection coefficient, $|R|^2$, is less than unity [5] despite the fact that the beam energy is less than the height of the barrier. Moreover, the missing electrons penetrating the potential step make the negative energy continuum (vacuum) negatively charged. These two un-expected and paradoxical results were first reported 80 years ago by O. Klein [2]. The traditional resolution of this paradox is shown as Fig. 5 [6]. It is interpreted as follows: at the barrier, pair production takes place in which electrons of flux $|T|^2$ are reflected to the left and an equal flux of anti-electrons are transmitted to the right into the barrier. Thus, overall reflection of electrons to the left becomes total since $|T|^2 + |R|^2 = 1$. Nonetheless, the vacuum becomes positively charged. Moreover, pair production means that one had to use tools and resort to concepts outside one-particle relativistic quantum mechanics where the original problem was presented.

Now, we present our approach to the resolution of the paradox, which is carried out entirely within relativistic quantum mechanics. We do that by replacing the *physical* pair production process with *virtual* negative energy anti-particle incidence under the barrier. This is a plane wave solution of virtual negative energy anti-particles traveling to the left under the barrier and incident on it causing positive energy particle transmission to the left and negative energy anti-particle reflection to the right. CPT symmetry of the Dirac equation dictates that we incorporate it as mirror image of the traditional solution (see Fig. 6). It is also worth noting that the *mathematical* solution of the Dirac equation (1) in the Klein energy zone with the vector potential depicted in Fig. 1 (without the arrows) could be representing positive energy electrons incident on the barrier from left but it could as well be representing negative energy anti-electrons incident on the barrier from right. It turns out that the correct solution includes both scenarios while at the same time satisfying the physical boundary conditions. As explained above, the inclusion of the virtual incident beam is analogous to what is usually done in optics and electrostatics. These "mirror models" (optical or electrical) could be used in all problems with such physical configuration in which the *vector* potential has a gradient larger than *2m* and the negative energy region *extends to infinity*.

Therefore, we proceed by including the negative energy solution under the barrier as virtual anti-electrons incident from right. That is, for $x > 0$ it reads as follows

$$\chi(x) = \frac{1}{\sqrt{1+\beta^2}} \binom{i\beta}{1} e^{ipx} + \frac{C}{\sqrt{1+\beta^2}} \binom{-i\beta}{1} e^{-ipx}. \tag{7}$$

This is a negative image of (4), which represents a combination of two negative energy beams of anti-electrons within the potential step. One beam is incident on the barrier from right with unit amplitude and the other is reflected to the right with an amplitude *C*. For $x < 0$, the corresponding solution is

$$\chi(x) = \frac{D}{\sqrt{1+\alpha^2}} \binom{1}{-i\alpha} e^{-ikx}, \tag{8}$$

which represents a transmitted beam of positive energy electrons to the left with amplitude *D*. Continuity of the spinor wavefunction $\chi(x)$ at $x = 0$ gives

$$C = (\alpha\beta - 1)/(\alpha\beta + 1), \quad D = \frac{2i\beta}{\alpha\beta+1} \sqrt{\frac{1+\alpha^2}{1+\beta^2}}. \tag{9}$$

Thus, the reflection amplitude in this case is $\mathcal{R} = C$ and the transmission amplitude is $\mathcal{T} = D\sqrt{\frac{\alpha}{\beta}\frac{1+\beta^2}{1+\alpha^2}}$. Consequently, the observed overall reflection of electrons to the left becomes total since

$$|R^-|^2 \equiv |R|^2 + |\mathcal{T}|^2 = 1. \tag{10}$$

Likewise, the overall virtual reflection of anti-electrons to the right is also total,

$$|R^+|^2 \equiv |\mathcal{R}|^2 + |T|^2 = 1. \tag{11}$$

Figure 7, illustrates the complete process, which is summarized in Fig. 8. In electrostatics, where the image charges replicate the virtual incident negative energy solutions, it is worthwhile to reassert the following. In the electrostatic problem, the *fictitious* image charge introduced in the unphysical region will replace the *real* charges induced on the surface at the boundary. In analogy, the *virtual* incident negative energy waves we introduced under the potential step does replace the *real* pair creation process.

The above constitutes the first successful attempt at a resolution of the 80-year old paradox within one-particle relativistic quantum mechanics. It results in *total reflection without charging the vacuum*. One can also show that the current density, *J(x)*, vanishes at the boundary $x = 0$ where $J(x) = -i\psi(x)^\dagger \sigma_3 \sigma_1 \psi(x)$, $\sigma_3 = \begin{pmatrix} +1 & 0 \\ 0 & -1 \end{pmatrix}$ and $\sigma_1 = \begin{pmatrix} 0 & 1 \\ 1 & 0 \end{pmatrix}$. Consequently, an incident wave packet, which is sharply centered within the Klein energy zone ($+m < E < V - m$), will be totally reflected with *zero probability* of penetrating the potential barrier. However, if the wave packet is sharply centered within the energy range $V \pm m$, then again it will be totally reflected but with non-vanishing probability of barrier penetration. On the other hand, if the wave packet is sharply centered around an energy greater than $V+m$, then partial reflection and transmission will occur. Now, if the energy spectrum of the wave packet extends over all three energy zones, then a linear combination of these three scenarios will occur.

Finally, we like to note that the understanding, interpretation and earlier attempts at a resolution of the Klein paradox presented few challenges in theoretical physics that lead to significant contributions, which proved to be very enriching and fruitful [7]. Unlike early attempts to resolve the paradox [6], the present one is carried out entirely within one-particle relativistic quantum mechanics where the original paradox was presented. Moreover, the negative energy continuum (vacuum) remains neutral. We believe that the resolution of the paradox offered here might have as significant implications as the original paradox itself.

**Acknowledgements:** This work is sponsored by the Saudi Center for Theoretical Physics (SCTP). Partial support by King Fahd University of Petroleum and Minerals (KFUPM) is acknowledged.

**Figure Caption:**

**Fig. 1**: The physical configuration associated with the Klein paradox problem. The height of the potential step is $V > 2m$ and the energy is in the range $V - m > E > +m$. The positive (negative) energy continuum is the region with the lighter (darker) shade. The solid line represents the vector potential $V$ and the dashed lines represent $\pm m$ or $V \pm m$. The energy level is indicated by the dashed-dotted line.

**Fig. 2**: The square barrier problem with the height of the barrier being larger than $2m$. Electrons are represented by the outlined arrows (⇨) and anti-electrons are represented by the solid arrows (➔).

**Fig. 3**: Electrostatic analog model of the atomic process: The induced charges on the boundary in (a) correspond to the created particles (⇨) in Fig. 2, whereas the negative charges displaced to $+\infty$ (not shown) correspond to the created anti-particles (➔) in Fig. 2. The virtual image charge in (b) corresponds to the virtual negative energy incidence.

**Fig. 4**: Optical analog model of the atomic process: (a) The flat mirror located at $x = 0$ corresponds to the discontinuity of the vector potential. (b) The region behind the mirror to the right where the imaginary light source is located corresponds to the negative energy continuum.

**Fig. 5**: Representation of the traditional resolution of the Klein paradox where spontaneous pair production takes place at the barrier. Positive and negative energy states are created with an equal flux, $|T|^2$.

**Fig. 6**: The virtual negative energy incidence from within the barrier is included as mirror image of the traditional solution. It results in the transmission and reflection coefficients $|\mathcal{T}|^2$ and $|\mathcal{R}|^2$, respectively.

**Fig. 7**: Representation of the proposed resolution of the Klein paradox within one-particle relativistic quantum mechanics (no pair production).

**Fig. 8**: The complete solution of the Klein paradox: (1) total reflection of positive and negative energy waves, and (2) particle/anti-particle conservation leaving a neutral vacuum.

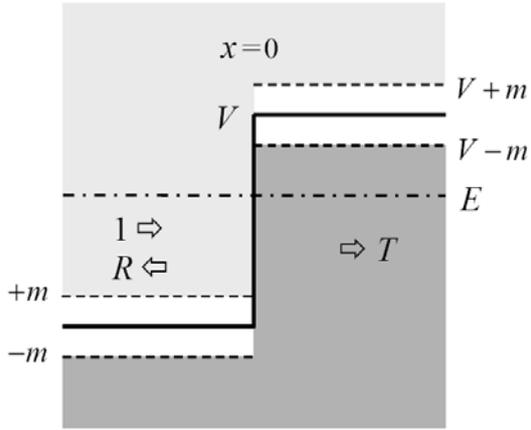 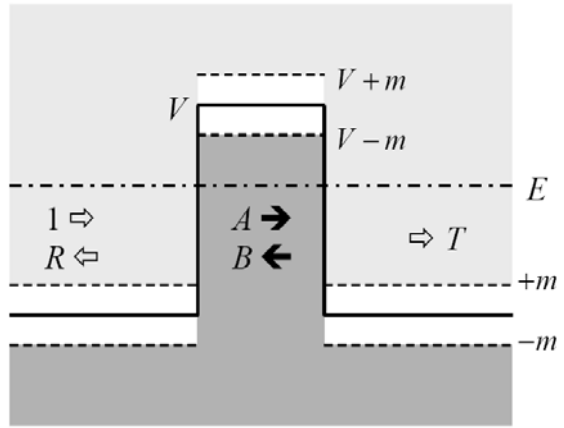

**Fig. 1**  **Fig. 2**

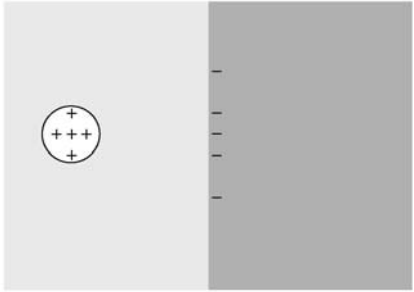 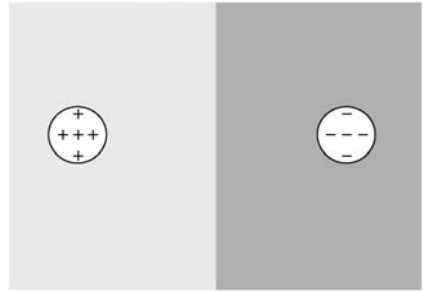

**Fig. 3a**  **Fig. 3b**

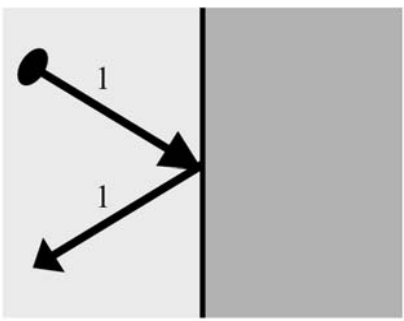 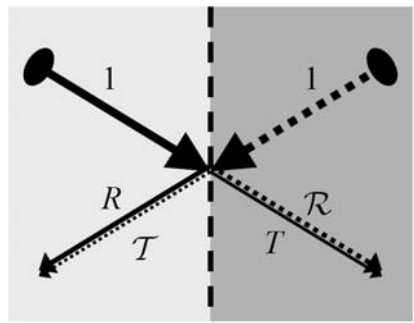

**Fig. 4a**  **Fig. 4b**

–8–

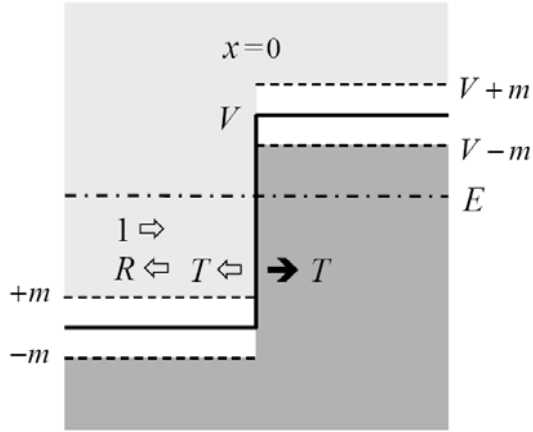
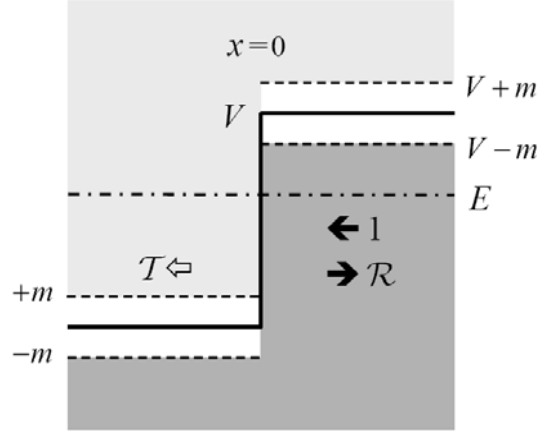

Fig. 5                                Fig. 6

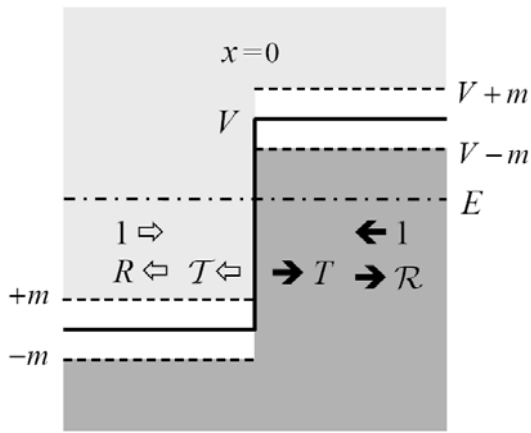
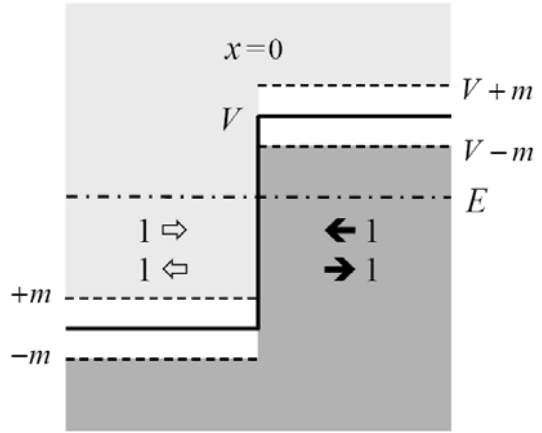

Fig. 7                                Fig. 8